\documentclass{article}
\usepackage{waspaa21,amsmath,graphicx,url,times}
\usepackage{amsfonts,amssymb}
\usepackage{color}
\usepackage{booktabs}

\usepackage{amsfonts}

\title{VQCPC-GAN: Variable-length Adversarial Audio Synthesis using Vector-Quantized Contrastive Predictive Coding}

\name{Javier Nistal,\sthanks{\scriptsize Nistal received funding from the European Union’s Horizon 2020 research and innovation programme under the Marie Skłodowska-Curie grant agreement No. 765068.}$^{,1,2}$
      Cyran Aouameur,$^{1}$
      Stefan Lattner,$^{1}$
      Gaël Richard$^{2}$
      }
\address{$^1$ Sony Computer Science Laboratory, 6 Rue Amyot, 75005 Paris, France\\   
         $^2$ LTCI, Télécom Paris, Institut Polytechnique de Paris, France\\
}

\begin{document}

\ninept
\maketitle

\begin{sloppy}

\begin{abstract}
  Influenced by the field of Computer Vision, Generative Adversarial Networks (GANs) are often adopted for the audio domain using fixed-size two-dimensional spectrogram representations as the ``image data''. However, in the (musical) audio domain, it is often desired to generate output of variable duration. This paper presents VQCPC-GAN, an adversarial framework for synthesizing variable-length audio by exploiting Vector-Quantized Contrastive Predictive Coding (VQCPC). A sequence of VQCPC tokens extracted from real audio data serves as conditional input to a GAN architecture, providing step-wise time-dependent features of the generated content. The input noise $z$ (characteristic in adversarial architectures) remains fixed over time, ensuring temporal consistency of global features. We evaluate the proposed model by comparing a diverse set of metrics against various strong baselines. Results show that, even though the baselines score best, VQCPC-GAN achieves comparable performance even when generating variable-length audio. Numerous sound examples are provided in the accompanying website,\footnote{\scriptsize{sonycslparis.github.io/vqcpc-gan.io}} and we release the code for reproducibility.\footnote{\scriptsize{github.com/SonyCSLParis/vqcpc-gan}}
\end{abstract}

\begin{keywords}
Generative Adversarial Networks, Audio Synthesis, Vector-Quantized Contrastive Predictive Coding
\end{keywords}

\section{Introduction}
\label{sec:intro}
\vspace{-0.2cm}
In recent years, Generative Adversarial Networks (GANs) \cite{Goodfellow2013} have shown outstanding results in image and audio synthesis tasks \cite{Karras2017, DBLP:conf/cvpr/KarrasLAHLA20, gansynth, nistal2, DBLP:conf/iclr/BinkowskiDDCECC20}. As most (initial) studies on GANs focused on generating images, the resulting architectures are now often adopted for the musical audio domain, using fixed-size two-dimensional spectrogram representations as the ``image data''. However, while it is a natural choice to use data of fixed dimensionality in the visual domain, fixing the length of musical audio content in generation tasks poses a significant limitation. As a result, GANs are currently mainly used to generate short audio content in the musical audio domain, like single notes of a tonal instrument or single percussion samples \cite{gansynth, nistal2}.

When dealing with variable-length sequence generation, commonly utilized models are Transformer architectures \cite{DBLP:journals/corr/abs-2004-10120}, causal convolutional architectures (causal CNNs) \cite{wavenet}, and recurrent neural networks (RNNs) \cite{DBLP:conf/interspeech/FanQXS14}. 
However, those models suffer various problems like high computational cost (autoregressive), missing look-back capabilities (recurrent), and, typically, they cannot be parallelized at test time.
In contrast, GANs are relatively efficient in generating high-dimensional data, as the conditioning on a single noise vector determines the values of all output dimensions at once. Therefore, it seems reasonable to also adopt the GAN paradigm for generating variable-length musical audio content. It has been shown in text-to-speech translation \cite{DBLP:conf/iclr/BinkowskiDDCECC20} that GANs can be successful in generating coherent variable-length audio when conditioned on meaningful sequences of symbols (i.e., linguistic + pitch features), while the input noise $z$ accounts for the remaining variability.

We adopt a similar strategy by first learning sequences of symbolic audio descriptors, serving as conditional inputs to a GAN architecture. These descriptors are discrete tokens learned through self-supervised training, using Vector-Quantized Contrastive Predictive Coding (VQCPC) \cite{DBLP:journals/corr/abs-2004-10120}. 
In VQCPC, discrete representations are learned through contrastive learning, by confronting positive and negative examples. In contrast to reconstruction-based VQ-VAEs \cite{DBLP:conf/nips/OordVK17}, VQCPC allows to control to some extent which aspects of the (sequential) data are captured in the tokens, by carefully designing a negative sampling strategy, thus defining the so-called ``pretext'' task.
In this work, the tokens are trained to represent the \emph{temporal evolution} (i.e., the envelope) of single, pitched audios of different instruments. The proposed model is conditioned on such envelope feature sequences, on the noise vector $z$ (static, representing the ``instrument''), and on pitch information (static).
This approach of sequence generation with GANs using discrete tokens is promising for future, more elaborate applications. While in this work, we are simply up-sampling token sequences to generate longer sounds, one could also \emph{generate} plausible token sequences. Such a system could then be used to hold sounds for an arbitrary time in real-time performance with a MIDI input device. Also, token sequences could be generated conditioned on MIDI information, to represent the dynamics of a target instrument. The resulting system could then be used for naturalistic rendering of MIDI files. Furthermore, training tokens to also represent pitch information would result in a more general variable-length audio generation framework. To the best of our knowledge, this is the first work implementing a variable-length GAN for \emph{musical} audio.

The paper is organized as follows. First, in Sec.~\ref{sec:previous_work}, we summarize previous works on adversarial audio synthesis, time-series GANs, and contrastive predictive coding. In Sec.~\ref{sec:vqcpcgan} we describe in detail the proposed framework. Sec.~\ref{sec:method} describes the experiment setup. Next, in Sec.~\ref{sec:results}, we evaluate the proposed method and compare results with previous work and other baselines. Finally, in Sec.~\ref{sec:conclusion}, we draw some conclusions and discuss future directions.

\vspace{-0.3cm}
\section{Previous work}
\label{sec:previous_work}
\vspace{-0.3cm}
In the following, we review some of the most important works on adversarial audio synthesis, variable-length time-series generation using GANs, and contrastive learning of sequences. We pay special attention to those works focused on audio data.
\begin{itemize}
    \item \textbf{Adversarial audio synthesis}. Applications of GANs to audio synthesis have mainly focused on speech tasks \cite{DBLP:conf/iclr/BinkowskiDDCECC20, DBLP:conf/nips/KongKB20, melgan}. The first GAN to synthesize musical audio was WaveGAN \cite{wavegan}. Recently, GANSynth \cite{gansynth} surpassed WaveNet \cite{wavenet} baselines on the task of audio synthesis of musical notes using pitch as conditional information. Follow-up works applied similar architectures to drum sound synthesis \cite{nistal2, Drysdale2020ADVERSARIALSO}. DrumGAN \cite{nistal2} performs high-quality synthesis of a variety of drum sounds conditioned on continuous timbral features. Other works employed GANs for Mel-spectrogram inversion \cite{melgan}, audio domain adaptation \cite{Hosseini} or audio enhancement \cite{DBLP:conf/icassp/BiswasJ20}. In this work, we adapt the Wasserstein GAN (WGAN) described in previous works \cite{gansynth, nistal2} to a sequence generation scheme by conducting two major architectural changes that we present in Sec.~\ref{sec:vqcpcgan}.

    \item \textbf{Time-Series GAN}. Early approaches to adversarial time-series generation used RNNs for both the generator and discriminator's architecture \cite{DBLP:conf/aaai/YuZWY17, DBLP:journals/corr/Mogren16}. C-RNN-GAN \cite{DBLP:journals/corr/Mogren16} generates musical data recurrently, taking as input a noise vector and the previous step's generated data. 
    While this approach relied only on the binary adversarial feedback for learning, TimeGAN \cite{DBLP:conf/nips/YoonJS19} incorporated an autoregressive supervised loss to better capture step-wise temporal dynamics of the training data. To maintain global and local consistency, the generator is conditioned on static and sequential random vectors. Similarly, GAN-TTS \cite{DBLP:conf/iclr/BinkowskiDDCECC20} synthesizes variable-length speech by conditioning the generator on sequential linguistic and pitch features, as well as a global random vector and a speaker ID. We take inspiration from these works and condition the generator on \emph{static} and \emph{dynamic} prior information. The \emph{static} information is represented by a random vector and a pitch class, whereas the \emph{dynamic} information is captured by a sequence of discrete tokens learned through self-supervised training.

    \item \textbf{Contrastive Predictive Coding (CPC)} \cite{DBLP:journals/corr/abs-1807-03748} is a self-supervised  framework used to learn general features from an unlabeled dataset of sequences by contrasting \emph{positive} and \emph{negative} examples in a so-called pretext task. CPC has been actively studied for speech tasks \cite{DBLP:journals/corr/abs-1807-03748, DBLP:conf/interspeech/SchneiderBCA19, DBLP:conf/iclr/BaevskiSA20}, where it was shown to improve the performance of ASR systems when used as front-end in replacement of spectrograms \cite{DBLP:conf/interspeech/SchneiderBCA19}. Introducing a VQ bottleneck to the CPC improved the system's performance by discarding irrelevant information \cite{DBLP:conf/iclr/BaevskiSA20, DBLP:conf/interspeech/NiekerkNK20}. In contrast to previous works exploiting VQCPC for discriminative downstream tasks \cite{DBLP:journals/corr/abs-1807-03748, DBLP:conf/interspeech/SchneiderBCA19, DBLP:conf/iclr/BaevskiSA20}, recent approaches explore small codebook sizes to learn compact, discrete representations of symbolic music from which to generate variations of any music piece \cite{DBLP:journals/corr/abs-2004-10120}. We follow a similar strategy in this work and use VQCPC to condition a GAN on such discrete codes for synthesizing variable-length audio.
\end{itemize}

\vspace{-0.3cm}
\section{VQCPC-GAN}
\vspace{-0.2cm}
\label{sec:vqcpcgan}
In this section, we briefly describe Contrastive Predictive Coding (CPC) and a variant for discrete representations, the Vector-Quantized CPC (VQCPC) \cite{DBLP:journals/corr/abs-2004-10120}, as well as the two building blocks of VQCPC-GAN, the VQCPC encoder and the GAN.
\vspace{-0.2cm}
\subsection{Background}
\label{sec:vqcpc}
Contrastive Predictive Coding (CPC) is a self-supervised representation learning technique for extracting compact, low-dimensional sequences of latent codes from high-dimensional signals \cite{DBLP:journals/corr/abs-1807-03748}.  Given an input sequence $\pmb{x} = [x_1, . . . , x_L]$ with length $L$, 
 an encoder $f_{\mathrm{enc}}$ maps each element $x_i$ into a real-valued embedding vector $z_i = f_{\mathrm{enc}}(x_i) \in R^{d_z}$. Next, an autoregressive model $f_{\mathrm{ar}}$ summarizes past and present context of the embeddings $z_{\le t}$ into a single context vector $ h_t = f_{\mathrm{ar}}(z_{\le t}) \in R^{d_h}$.

The encoder and autoregressive model are trained to minimize the Information Noise Contrastive Estimation (InfoNCE) loss. Minimizing the InfoNCE loss is equivalent to maximizing the mutual information between the context vector $h_t$ and the encoding of future subsequences $z_{t+k} = f_{\mathrm{enc}}(x_{t+k}), \forall k \in [1, K]$, where K is the number of future predictions  \cite{DBLP:journals/corr/abs-1807-03748}. 
Formally, given an entry of the dataset $\pmb{x}$, the model has to identify the encoding obtained from the true $x_{t+k}$, so-called \emph{positive example}, from those obtained from a set of so-called \emph{negative examples}, drawn from a proposal distribution $p(x_{t+k})$. Defining $\mathcal{S}$ as the set containing $N - 1$ negative examples as well as the single positive example $x_{t+k}$, the InfoNCE loss is written as
\vspace{-0.1cm}

\begin{scriptsize}
\begin{equation}\label{eq:lossNCE}
    \mathcal{L}_{\mathrm{NCE}}(x_t) = - \sum_{k=1}^{K} \mathop{E}_\mathcal{S} \left[\log\frac{f_k(x_{t+k}, h_t)}{\sum_{s \in \mathcal{S}} f_k(s , h_t)}\right],
\end{equation}
\end{scriptsize}

where $\mathop{E[\cdot]}$ denotes expectation and $f_k(a, b) :=
\exp(f_{\mathrm{enc}}(a)^\intercal W_kb)$ is a simple log-bilinear model with the $W_k$ being $k$ trainable $d$ x $d$ matrices.


VQCPC \cite{DBLP:journals/corr/abs-2004-10120} introduces a Vector Quantization (VQ) \cite{DBLP:conf/nips/OordVK17} bottleneck on top of the CPC encoder to obtain a discrete latent representation of the subsequences:
\begin{scriptsize}
\begin{equation}
    \mathrm{VQCPC}(x_i) := z^q(f_{\mathrm{enc}}(x_i)) \in \mathcal{C},
\end{equation}
\end{scriptsize}
where the \emph{codebook} $\mathcal{C}$ is a set of $C$ centroids partitioning the embedding space $R^{d_c}$.

\vspace{-.2cm}
\subsection{The VQCPC encoder}\label{sec:vqcpc_enc}
The VQCPC encoder $f_{\mathrm{enc}}$ is a stack of 4 convolutional blocks operating frame-by-frame. Each block is composed of a 1D CNN (in time) with a kernel size of 1 and number of channels (512, 512, 256, $d_z$) respectively for each block in the stack. Each CNN, except the last one, is followed by a ReLU. As opposed to \cite{simCLR}, there is no projection head.
VQ is trained with a squared $\mathcal{L}_2$ loss with a commitment component \cite{DBLP:conf/nips/OordVK17}. We choose a codebook $\mathcal{C}$ containing $C=16$ centroids, and where $d_c = d_z = 32$.
The codebook size is chosen relatively small, enforcing an information bottleneck that only lets through the most salient information needed to discriminate between positive and negative examples \cite{DBLP:journals/corr/abs-2004-10120}.
The autoregressive model $f_{\mathrm{ar}}$ is a 2-layer GRU with a hidden size of 256 and an output size of 512, and we use its output at timestep $t$ as the context vector $h_t$ to predict $K=5$ timesteps into the future.
The overall training objective is the VQ 
and the 
InfoNCE loss (\ref{eq:lossNCE}).

As mentioned earlier, an important choice in contrastive learning is the design of the negative sampling strategy, as this controls what features are represented by the encoding. Usually, the proposal distribution for the negative samples is chosen to be uniform over the training set \cite{DBLP:journals/corr/abs-2004-10120, DBLP:journals/corr/abs-1807-03748, DBLP:journals/corr/abs-1905-09272}. However, in this work, we sample 16 negative examples in an intra-sequence fashion: given an audio excerpt $\pmb{x}$, the negative examples 
are all drawn from a uniform distribution over $\pmb{x}$ (i.e., the same audio excerpt). 
This intra-sequence sampling forces the network to encode only information which varies \emph{within a sample} (i.e., dynamic information such as onset, offset, loudness change, vibrato, tremolo, etc.), while ignoring static information like instrument type and pitch. This shows that VQCPC provides a convenient way to control what should be represented by the discrete representations. In this work, the remaining information (instrument type and pitch) is represented by the GAN's input noise and the explicit pitch conditioning.

\subsection{The GAN architecture}\label{sec:arch}

\begin{figure}[t]
    \centering

    \includegraphics[scale=0.6, width=0.92\columnwidth]{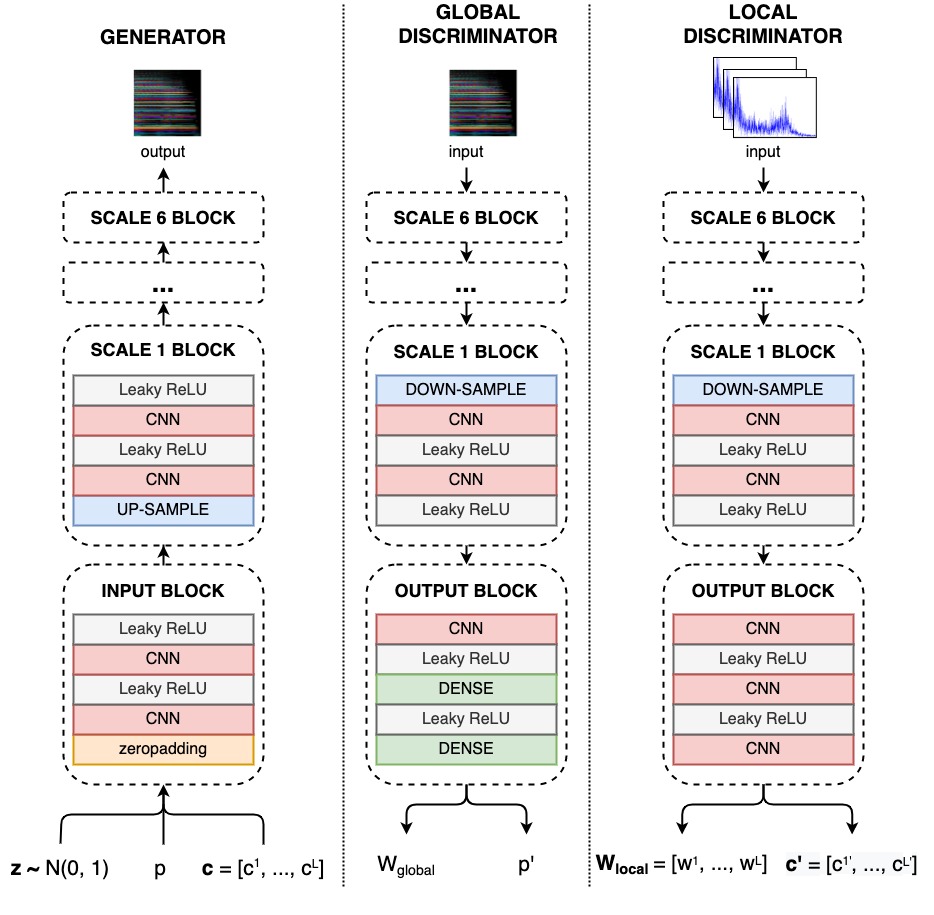}
    \vspace{-.4cm}
    \caption{Proposed architecture for VQCPC-GAN (see Sec.~\ref{sec:arch}).}
    \vspace{-.4cm}
    \label{fig:arch}
\end{figure}

    The proposed WGAN is inherited from DrumGAN \cite{nistal2}. We adapt the architecture to a sequential scheme by conducting two major changes. First, the input tensor to the generator $G$ is a sequence containing \emph{static} and \emph{dynamic} information. The \emph{static} information refers to the global context and accounts for the pitch class, a one-hot vector $p \in \{0,1\}^{26}$ with $26$ possible pitch values, as well as a noise vector $z \sim N(0, 1) \in \mathbb{R}^{128}$ sampled from a standard normal distribution with zero mean and unit variance $N(0, 1)$. The \emph{dynamic} information provides local frame-level context and is composed of a sequence of discrete, one-hot vectors $\textbf{c} = [c^1, ..., c^L]$ where $c^l \in \{0,1\}^{16}$ and $L$ is the number of frames in the sequence. The tensor $\textbf{c}$ identifies a sequence of spectrogram clusters obtained by encoding real audio using VQCPC (see Sec.~\ref{sec:vqcpc}). At training time, $L$ is set to $32$ frames, which corresponds to approximately $1$ second of audio given the pre-processing parameters (see Sec.~\ref{sec:data}). The \emph{static} vectors $p$ and $z$ are repeated across the sequence dimension $L$ of the \emph{dynamic} information $\textbf{c}$, resulting in a tensor $\textbf{v} \in \mathbb{R}^{L\times160}$. This tensor is unsqueezed, reshaped to ($160 \times 1 \times L$) and fed through a stack of convolutional and nearest-neighbour up-sampling blocks to generate the output signal $x=G_\theta(z; \textbf{c}; p)$. In order to turn the input tensor into a spectrogram-like convolutional input, it is first zero-padded in the frequency dimension.  As depicted in Fig.~\ref{fig:arch}, the generator's input block performs this zero-padding followed by two convolutional layers with ReLU non-linearity. Each scale block is composed of one nearest-neighbour up-sampling step at the input and two convolutional layers with filters of size $(3, 3)$. The number of feature maps decreases from low to high resolution as \{512, 256, 256, 256, 256, 128\}. We use Leaky ReLUs as activation functions and apply pixel normalization. 

The second major change is the use of two discriminators (see Fig.~\ref{fig:arch}). A local discriminator $D_l$, implemented in a fully convolutional manner, estimates $W_{local}$ which is the Wasserstein distance \cite{Gulrajani2017} between real and generated distributions at a frame-level (i.e. using batches of frames instead of batches of full spectrograms). Additionally, to encourage $G$ to consider the conditional sequence of VQCPC tokens, $D_l$ performs an auxiliary classification task where each input spectrogram frame is assigned to a VQCPC token $c_l$. We add an additional cross-entropy loss term for $D_l$'s objective \cite{OdenaOS17}. A global discriminator $D_g$ with two dense layers in its output block estimates $W_{global}$ over complete sequences of $L=32$ spectrogram frames and predicts the pitch class. As in $D_l$, we add an auxiliary cross-entropy loss term to $D_g$'s objective function for the pitch classification task \cite{OdenaOS17}.

\vspace{-.2cm}
\subsection{Training Procedure}
As in \cite{nistal2}, training follows Progressive Growing of GANs (ProGAN) \cite{Karras2017} where the architecture is built dynamically during training. The process is divided into scale iterations wherein each scale a new block is added to $G$ and $D$ (see Fig.~\ref{fig:arch}). The models are trained 200k iterations per scale using batches of 30, 30, 20, 20, 12, and 12, respectively. We employ Adam as the optimization method with a learning rate of 0.001 in both networks.

\vspace{-.3cm}
\section{Experiments}
\vspace{-.1cm}
\label{sec:method}
In this work, we employ a VQCPC encoder (see Sec. \ref{sec:vqcpc}) to learn discrete sequences of high-level features from a dataset of tonal sounds (see Sec. \ref{sec:data}). As described in Sec. \ref{sec:arch}, we condition a GAN on such discrete sequential representations in order to perform audio synthesis. Variable-length audio is achieved by up/down-sampling, respectively for longer or shorter sounds, of the conditional VQCPC sequence. In the following, we present the training dataset, the evaluation metrics and the baselines.
\vspace{-.3cm}
\subsection{Dataset} \label{sec:data}
We employ a subset of audio excerpts obtained from the NSynth dataset \cite{wavenetae}. The subset contains over 50k single-note audios played by 11 instrument families. Among the available annotations, we make use of the pitch class. The (monophonic, 16kHz) audio clips are trimmed to 1 second. We only consider samples with a pitch in the MIDI range from 44 to 70 (103.83 - 466.16 Hz). For the evaluation, we use a 90/10\% split of the data.

Previous works indicated that magnitude and Instantaneous Frequency of the Short-time Fourier Transform (STFT) are well-suited for structuring tonal audio information \cite{nistal1, gansynth, DBLP:journals/corr/abs-2104-07519}. We, therefore, preprocess the data with this transform using an FFT size of 2048 bins and an overlap of 75\%. For the VQCPC encoder (see~\ref{sec:vqcpc_enc}), we rely on the Constant-Q Transform (CQT) spanning 6 octaves with 24 bins per octave. We use a hop-length of 512 samples, for the output token sequence to match the temporal resolution of the data used to train the GAN.

\vspace{-.3cm}
\subsection{Evaluation}\label{sec:eval}
\vspace{-.1cm}
Similarly to previous works on adversarial audio synthesis \cite{gansynth, nistal1, nistal2, DBLP:journals/corr/abs-2103-07390}, we evaluate our models using the following metrics:


\begin{itemize}
    \vspace{-0.2cm}
    \item The \textbf{Inception Score (IS)} \cite{SalimansGZCRCC16} is defined as the mean KL divergence between the conditional class probabilities \(p(y|x)\), and the marginal distribution \(p(y)\) using the class predictions of an Inception classifier. IS penalizes models (low IS) that generate examples that cannot be easily classified into a single class with high confidence as well as models whose examples belong to only a few of all possible classes. We train our own inception model to classify  pitch and instrument on the NSynth dataset yielding two metrics: the Pitch Inception Score (PIS) and Instrument Inception Score (IIS) \cite{nistal1}.
    \vspace{-0.1cm}
    \item The \textbf{Kernel Inception Distance (KID)} \cite{BinkowskiSAG18} is defined as the squared Maximum Mean Discrepancy (MMD) between embeddings of real and generated data in a pre-trained Inception-like classifier.
     The MMD measures dissimilarity between real and generated distributions therefore the lower the better. We compute the KID with the same Inception model as in IS.
    \vspace{-0.1cm}
    \item The \textbf{Fréchet Audio Distance (FAD)} \cite{fad} is computed as the difference between multivariate Gaussian distributions fitted to the output embedding of a pre-trained VGG-like model.
    A lower FAD indicates a smaller distance between real and generated distributions. We employ Google's FAD implementation.\footnote{\scriptsize \url{https://github.com/google-research/google-research/tree/master/frechet_audio_distance}}

\end{itemize}
\vspace{-.4cm}
\subsection{Baselines}
\vspace{-.1cm}
We compare the metrics described in Sec.~\ref{sec:eval}  with a few baselines and include results scored by \emph{real data} to delimit the range of each metric. Specifically, we compare with GANSynth \cite{gansynth}, obtained from Google Magenta's github,\footnote{\scriptsize \url{https://github.com/magenta/magenta/tree/master/magenta/models/gansynth}} and two baselines that we train using the same architecture of VQCPC-GAN but removing the sequence generation scheme, i.e., without VQCPC conditioning nor the local $D$. We train the two baseline models, WGAN$_{1s}$ and WGAN$_{4s}$, on $1s$ and $4s$-long audio excerpts, respectively, whereas GANSynth is originally trained on $4s$ audio excerpts. 
As mentioned early on in this section, we condition VQCPC-GAN on varying-length VQCPC sequences in order to generate audio with different duration. To do so, we just up/down-sample the VQCPC sequence accordingly to obtain the desired number of output frames. In particular for these experiments, we take the original VQCPC sequences of length 32 (i.e., $1s$-long) and perform nearest-neighbour up-sampling by a factor of 4 to obtain 128 tokens (i.e., $4s$-long).

\vspace{-0.3cm}
\section{Results and discussion}
\label{sec:results}
\vspace{-0.23cm}
In this section, we present the results from the evaluation metrics described in Sec.~\ref{sec:eval}. We informally validate these quantitative results by listening to the generated content and sharing our assessment.

\vspace{-0.23cm}
\subsection{Quantitative results}\label{sec:res1}
\vspace{-0.13cm}
Table \ref{tab:metrics} presents the metrics scored by our proposed VQCPC-GAN and the baselines. Overall, our WGANs score closest to those of \emph{real data} in most metrics, or even better in the case of the PIS. GANSynth follows closely and VQCPC-GAN obtains slightly worse results. VQCPC-GAN performs particularly good in terms of PIS, which suggests that the generated examples have an identifiable pitch content and that the distribution of pitch classes follows that of the training data. This is not surprising given that the model has explicit pitch conditioning, making it trivial to learn the specific mapping between the pitch class and the respective tonal content. Conversely, results are worse in the case of IIS, suggesting that the model failed to capture the timbre diversity existent in the dataset and that generated sounds cannot be reliably classified into one of all possible instrument types (i.e. mode failure). Turning now our attention to the KID, VQCPC-GAN scores results very similar to GANSynth and slightly worse than WGAN$_{1s}$. A low KID indicates that the Inception embeddings are similarly distributed for real and generated data. Our Inception classifier is trained on several discriminative tasks of specific timbral attributes, including pitch and instrument classification. Therefore, we can infer that similarities in such embedding space indicate shared timbral and tonal characteristics, from a statistical point of view, between real and generated audio data. This trend is not as evident in the case of the FAD, where VQCPC-GAN obtains considerably worse results than the baselines, particularly in the case of WGAN$_{1s}$. This could indicate the existence of artefacts as FAD was found to correlate well with several artificial distortions \cite{fad}.

To wrap up: despite the architectural changes introduced for sequential generation, VQCPC-GAN exhibits results comparable to GANSynth, the SOTA on adversarial audio synthesis of tonal sounds, as well as two strong baselines WGAN$_{1,4s}$ trained on 1 and 4-second long audio respectively. Notably, our WGAN$_{4s}$ baseline scores better results than GANSynth in all metrics. In the following section, we informally validate these quantitative results by sharing our assessment when listening to generated audio material.

\vspace{-0.32cm}
\subsection{Informal listening}
\vspace{-0.2cm}
The accompanying website contains audio examples generated under different settings (e.g. latent interpolations, pitch scales, generation from MIDI files) and different duration (0.5, 1, 2 and 4 seconds). Synthesis of variable-length audio is achieved by up/down-sampling of the conditional VQCPC sequence. Overall, we find the results discussed in Sec.~\ref{sec:res1} to align well with our perception. The range of instruments is narrow, and only a few from the most homogeneous and populated classes in the dataset can be identified (e.g., mallet, guitar, violin), hence the low IIS. In the \emph{pitch scale} examples, we can perceive that the pitch content responds nicely to the conditional signal, and it is consistent across the generation time span, which explains the higher PIS. Although we eventually obtain some artifacts when using certain VQCPC token combinations as conditional input, the overall quality is acceptable. This is aligned with having a low FAD but a KID comparable to the baselines.
\begin{scriptsize}
\begin{table}
    \centering
    \scriptsize
\begin{tabular}[]{c c c c c c c c c}
       & \multicolumn{2}{c}{IIS $\uparrow$} & \multicolumn{2}{c}{PIS $\uparrow$} & \multicolumn{2}{c}{KID$^a$ $\downarrow$} & \multicolumn{2}{c}{FAD $\downarrow$}\\
     \toprule
      duration (s) & 1 & 4 & 1 & 4 & 1 & 4 & 1 & 4 \\
     \midrule
     \emph{real data} & 6.3 & 4.5 & 17.9 & 18.0 & 6.7 & 6.6 & 0.0 & 0.0 \\
     \midrule
     GANSynth \cite{gansynth} & - & 4.1 & - & 19.7 & - &7.0 & - & 2.1\\
     WGAN$_{1s}$ &\textbf{4.5}& - &\textbf{19.0}&-&\textbf{6.8}&-& \textbf{0.8} & - \\
     WGAN$_{4s}$ &-&\textbf{4.5}&-&\textbf{20.1}&-&\textbf{6.9}&-&\textbf{1.0} \\
     VQCPC-GAN & 3.0 & 2.9 & 18.5 & 17.2 & 7.3 & 7.1 & 5.6 & 5.4 \\
     \bottomrule
     \multicolumn{9}{l}{$^{\mathrm{a}}$\(\times{10^{-4}}\)}
     \label{tab:metrics}
\end{tabular}

\vspace{-.3cm}
\caption{\footnotesize IIS, PIS, KID, and FAD (Sec.~\ref{sec:eval}), scored by VQCPC-GAN and baselines. The metrics are computed over $25$k samples.}
\vspace{-.5cm}
\end{table}
\end{scriptsize}

\vspace{-0.3cm}
\section{Conclusion}
\label{sec:conclusion}
\vspace{-0.2cm}
In this work, we presented VQCPC-GAN, an adversarial model capable of performing variable-length sound synthesis of tonal sounds. We adapted the WGAN architecture found in previous works \cite{gansynth, nistal2} to a sequential setting by conducting two major architectural changes. First, we condition $G$ on \emph{dynamic} and \emph{static} information captured, respectively, by a sequence of discrete tokens learned through VQCPC, and a global noise \emph{z}. Additionally, we introduce a secondary fully-convolutional $D$ that discriminates between real and fake data distributions at a frame level and predicts the VQCPC token associated with each frame. Results showed that VQCPC-GAN can generate variable-length sounds with controllable pitch content while still exhibiting results comparable to previous works generating audio with fixed-duration. We provide audio examples in the accompanying website. As future work, we plan on investigating hierarchical VQCPC tokens to condition the GAN on longer-term, compact representations of audio signals.

\vspace{-0.3cm}
\section{ACKNOWLEDGMENTS}
\label{sec:ack}
\vspace{-0.3cm}
 The authors thank Gaëtan Hadjeres for his guidance and support.

\bibliographystyle{IEEEtran}
\bibliography{refs21}

\end{sloppy}
\end{document}